\documentstyle[12pt]{article}

\newcommand{\gsim}{\raisebox{-4pt}{$\,\stackrel{\textstyle >}{\sim}\,$}}
\newcommand{\fpg}[1]{F_{\pi\gamma}\left(#1\right)}
\newcommand{\pg}{$\pi\gamma$ }
\newcommand{\pgff}{\pg form factor}
\newcommand{\pgtff}{\pg transition form factor}
\newcommand{\da}{distribution amplitude}
\newcommand{\peff}{pion's electromagnetic form factor}
\newcommand{\pff}{pion form factor}
\newcommand{\bc}{\begin{center}}
\newcommand{\ec}{\end{center}}
\newcommand{\beq}{\begin{equation}}
\newcommand{\eeq}{\end{equation}}

\newcommand{\beqa}{\begin{eqnarray}}
\newcommand{\eeqa}{\end{eqnarray}}

\begin{document}

\begin{titlepage}
\hspace{10.5cm}WU-B~96-19

\hspace{10.5cm}April 1996\\[2em]
\begin{center}
\Large{THE $\pi\gamma$ TRANSITION FORM FACTOR AND THE PION WAVE FUNCTION}\\[2em]
P.~Kroll\footnote{kroll@theorie.physik.uni-wuppertal.de}, 
M.~Raulfs\footnote{Supported by the Deutsche Forschungsgemeinschaft.}\\
Fachbereich Physik, Universit\"at Wuppertal,\\
D-42097 Wuppertal, Germany
\end{center}
\end{titlepage}
\setcounter{footnote}{0}
\begin{titlepage}
\hspace{10.5cm}WU-B~96-19

\hspace{10.5cm}October 1995\\[2em]
\begin{center}
\Large{THE $\pi\gamma$ TRANSITION FORM FACTOR AND THE PION WAVE FUNCTION}\\[2em]
P.~Kroll\footnote{kroll@theorie.physik.uni-wuppertal.de}, 
M.~Raulfs\footnote{Supported by the Deutsche Forschungsgemeinschaft.}\\
Fachbereich Physik, Universit\"at Wuppertal,\\
D-42097 Wuppertal, Germany
\end{center}
\begin{abstract}
The pion wave function is discussed in the light of the recent CLEO data on
the \pgtff. It turns out that the wave function is close to the asymptotic 
form whereas wave functions strongly concentrated in the end-point regions are 
disfavoured. Consequences for other exclusive quantities, as for 
instance the \peff, are also discussed.
\end{abstract}
\end{titlepage}

The theoretical description of large momentum transfer exclusive reactions is
based on a factorization of long- and short-distance physics. The latter 
physics is contained in the so-called hard scattering amplitude to be 
calculated within perturbation theory. Universal, process-independent (light 
cone) wave functions interpolating between hadronic and partonic degrees of 
freedom, comprise the long-distance physics. The wave functions are not 
calculable with sufficient degree of accuracy at present. However, for the pion
which is the hadron of interest in this letter, the valence Fock state wave
function is theoretically rather well constrained. The main uncertainty of the 
wave function lies in the $x$-dependent part of it, the so-called distribution
amplitude $\phi(x)$. $x$ denotes the usual momentum fraction the valence quark 
carries.

Previous studies of the \peff\ as well as other large momentum transfer 
exclusive reactions, as for instance $\gamma\gamma\to\pi\pi$, seemed to indicate 
that the pion \da\ is much broader than the so-called asymptotic one, 
$\phi_{AS}(x)=6x(1-x)$. Chernyak and Zhitnitsky \cite{CheZhi:82} proposed such 
a broad \da\ which is strongly end-point concentrated and leads to a leading 
twist contribution to the \peff\ in apparently fair agreement with the 
admittedly poor data \cite{Beb:76}. This result is, however, obtained at the 
expense of the dominance of contributions from the end-point regions, $x\to 0$ 
or $1$, where the use of perturbative QCD is unjustified as has been pointed 
out by several authors \cite{IsgLle:89,Rad:91}. Now the prevailing opinion is 
that the \peff\ is controlled by soft physics (e.g. the overlap of the initial 
and final state wave functions, occasionally termed the Feynman contribution) 
for momentum transfer less than about $10$ GeV$^2$ 
\cite{IsgLle:89,JaKro:93,KisWan:93,Dor:95}.

There is another exclusive quantity namely the \pgtff\ which, for experimental
and theoretical reasons, allows a more severe test of our knowledge of the
pion wave function than the \peff. Recently the \pgff\ has been measured in 
the momentum transfer region from $2$ to $8$ GeV$^2$ \cite{Sav:95} with rather
high precision. Together with previous CELLO measurement \cite{Beh:91} we now 
have at our disposal much better data above $1$ GeV$^2$ for the \pgtff\ than 
for the \pff. From the theoretical point of view the analysis of the \pgtff\ 
is much simpler than that of the \pff: It is, to lowest order, a QED process, 
QCD only provides corrections of the order of $10$\% in the 
momentum transfer region of interest \cite{bra:83}. The difficulties with the end-point 
regions where the gluon virtuality becomes small, do not occur. Higher Fock 
state contributions, suppressed by powers of $\alpha_s/Q^2$, are expected to 
be small (see the discussion in \cite{Gor:89}). Moreover, and in contrast to 
the \pff, the Feynman contribution, which may arise through vector meson 
dominance, is presumably very small due to a helicity mismatch.

The purpose of this letter is to extract information on the pion wave function 
from a perturbative analysis of the \pgff. It will turn out that the new CLEO 
data \cite{Sav:95} allow a fairly precise determination of that wave function. 
It will also be argued that this wave function leads to a consistent 
description of the \peff\ and its structure function. It should be noted that 
this letter is an update of previous work \cite{JaKroRau:94}. The very 
important CLEO data \cite{Sav:95} were not yet available in 
\cite{JaKroRau:94}. We also point out that the $\pi\gamma^*$ transition form 
factor is investigated in \cite{Ong:95}.

Let us begin with the parameterization of the soft valence Fock state wave
function, i.e. the full wave function with the perturbative tail removed from
it. This is the object required in a perturbative calculation 
\cite{BroHuaLe:83}. Following \cite{JaKro:93} the soft wave function is 
written as
\beq
\hat\Psi_0\left(x,{\bf b},\mu_F\right)=\frac{f_\pi}{2\sqrt 6}\,\phi(x,\mu_F)\,
\hat\Sigma\left(\sqrt{x(1-x)}{\bf b}\right).
\label{wfpareq}
\eeq
$\mu_F$ is the scale at which soft and hard physics factorize 
\cite{BroHuaLe:83,BroLe:80,BoSte:89}. The wave function is subject to the 
auxiliary conditions
\beq
\hat\Sigma(0)=4\pi,\hspace{3cm}\int_0^1dx\,\phi(x,\mu_F)=1.
\eeq
$\bf b$ is the quark-antiquark separation in the transverse configuration
space and is canonically conjugated to the usual transverse momentum
$\bf k_\perp$. It is advantageous to work in the transverse configuration 
space because the Sudakov factor, to be discussed later, is only derived in 
that space (see \cite{BoSte:89}). The parameterization (\ref{wfpareq})
automatically satisfies the constraint from the process $\pi^+\to\mu^+\nu_\mu$
\cite{BroHuaLe:83} which relates the wave function at the `origin of the
configuration space` to the pion decay constant $f_\pi(=130.7\mbox{ MeV})$.
On the assumption of duality properties, one can derive constraints on $x$ 
and $k_\perp$ moments\footnote{
There is a subtlety: The kinematical transverse momentum of the parton is not
the same object as $k_\perp$ defined through the moments. In this letter 
we will assume that both are one and the same variable. This assumption 
corresponds to summing up soft gluon corrections, i.e. to higher twist 
contributions.}
 of the wave function (e.g. $\langle x^n\rangle=\int dx\,x^n\,\phi(x)$) within
the operator product expansion framework \cite{ChiZhi:95}. These constraints
can be combined into the following conditions on the momentum space wave
function:

\begin{tabular}{rl}
i)&The distribution amplitude has simple zeroes at $x\to0,1$.\\
ii)&The $\bf k_\perp$-dependence of the wave function $\Psi_0
\left(x,{\bf k_\perp}\right)$ comes exclusively\\
&in the combination $k_\perp^2/x(1-x)$ at $x\to0,1$.\\
iii)&At large $k_\perp$ the wave function falls off faster than any power of 
$k_\perp$.\\
\end{tabular}
\\[1em]

The simplest function matching the conditions ii) and iii) is the Gaussian
\beq
\hat\Sigma\left(\sqrt{x(1-x)}\,{\bf b}\right)=4\pi\exp\left[-\frac{x(1-x)\,b^2}{4a^2}
\right]
\label{gausskteq}
\eeq
where $a$ is the transverse size parameter. This Gaussian will be used 
subsequently.

The distribution amplitude is subject to evolution and can be expanded over 
Gegenbauer polynomials $C_n^{3/2}$, the eigenfunctions of the (leading
order) evolution equation for mesons
\beq
\phi(x,\mu_F)=\phi_{AS}(x)\left[1+\sum^\infty_{n=2,4,...}B_n\left(
\frac{\alpha_s\left(\mu_F\right)}{\alpha_s\left(\mu_0\right)}\right)
^{\gamma_n}\,C_n^{3/2}(2x-1)\right].
\label{evoleq}
\eeq
$\alpha_s$ is the strong coupling constant and $\mu_0$ is a typical hadronic 
scale for which we choose $0.5$ GeV throughout. Charge conjugation invariance 
requires the odd $n$ expansion coefficients $B_n$ to vanish. Since the 
$\gamma_n$ are positive fractional numbers increasing with $n$ (e.g. 
$\gamma_2=50/81$) any distribution amplitude evolves into $\phi_{AS}$ 
asymptotically, i.e. for $\ln(Q/\mu_0)\to\infty$; higher order terms in 
(\ref{evoleq}) are gradually suppressed. The asymptotic distribution
amplitude itself shows no evolution. This property of $\phi_{AS}$ 
no more holds if evolution is treated in next-to-leading order
\cite{mue:95}. As we are going to show below the 
asymptotic \da\ in combination with the Gaussian (\ref{gausskteq}) provides 
very good results for the \pgff\ and also leads to a consistent description of 
other exclusive reactions involving pions. In order to quantify the amount of 
deviations from the asymptotic \da\ still allowed by the CLEO data, 
contributions from the second Gegenbauer polynomial will be permitted in the 
analysis and limits for the strength of the coefficient $B_2$ will be 
extracted. For the purpose of comparison the \pgff\ will be evaluated with the 
Chernyak-Zhitnitsky (CZ) \da\ \cite{CheZhi:82} which is defined by $B_2=2/3$ 
and $B_n=0$ for $n>2$.

For a given \da\ there is only one free parameter in the wave function, namely 
the transverse size parameter $a$. It can be fixed by using a constraint 
derived from the process $\pi^0\to\gamma\gamma$ \cite{BroHuaLe:83}:
\beq
\int dx\,d^2{\bf b}\,\hat\Psi_0(x,{\bf b})=\frac{\sqrt 6}{f_\pi}.
\label{p0ggeq}
\eeq
Although not at the same level of rigour as the other constraints (because 
of approximations made in the case when one photon couples `inside` the pion 
wave function) (\ref{p0ggeq}) still provides a value for the parameter $a$ 
which comes up to our expectations for the transverse size of the pion. In 
particular for the asymptotic wave function one finds a value of $861$ MeV for 
$a$, which corresponds to a value of $367$ MeV for the root mean square 
transverse momentum. The probability of the valence Fock state amounts to 
$0.25$ in this case.

Now the wave function is fully specified and we can turn to the calculation of 
the \pgff. Inspection of the \pgff\ data \cite{Sav:95,Beh:91} (see Fig.~1) 
reveals a $Q^2$ dependence somewhat stronger than predicted by dimensional counting. 
Hence, higher twist contributions do not seem to be negligible in the momentum transfer 
region from $1$ to $8$ GeV$^2$. The modified perturbative approach proposed by
Sterman and collaborators \cite{BoSte:89,LiSte:92} allows to calculate some 
power corrections to the leading twist term. In that approach the transverse 
momentum dependence of the hard scattering amplitude is retained and Sudakov 
suppressions are taken into account in contrast to the standard approach 
\cite{BroLe:80}. Applications of the modified perturbative approach to the 
pion's and nucleon's electromagnetic form factors 
\cite{JaKro:93,LiSte:92,BoJaKro:94} revealed that the perturbative 
contributions to these form factors can self-consistently (in the sense that 
the bulk of the contributions is accumulated in regions where the strong 
coupling $\alpha_s$ is sufficiently small) be calculated. It turned out, 
however, that the perturbative contributions are too small as compared with 
the data. Therefore, in the experimentally accessible range of momentum 
transfer, these form factors are controlled by soft physics. Higher order
perturbative corrections and/or higher Fock state contributions seem too small 
in order to account for the large discrepancies between the lowest order 
perturbative contributions and the data for the elastic form factors
(see \cite{fie:81} for the pion case).

For the reasons discussed in the introduction the \pgtff\ represents an
exceptional case for which we can expect perturbation theory to work for $Q^2$ 
larger than about $1$ or $2$ GeV$^2$. Adapting the modified perturbative 
approach \cite{BoSte:89,LiSte:92} to the case of
\pg\ transitions, we write the corresponding form factor as
\beq
\fpg{Q^2}=\int dx\,\frac{d^2{\bf b}}{4\pi}
\hat\Psi_0\left(x,-{\bf b},\mu_F\right)\,
\hat T_H\left(x,{\bf b},Q\right)\,\exp\left[-S\left(x,b,Q\right)\right].
\label{fpgeq}
\eeq
This convolution formula can formally be derived using the methods described 
in detail by Botts and Sterman \cite{BoSte:89}. $\hat T_H$ is the Fourier 
transform of the momentum space hard scattering amplitude to be calculated, to 
lowest order, from the Feynman graphs shown in Fig. 2. It reads
\beq
\hat T_H\left(x,{\bf b},Q\right)=\frac{2}{\sqrt 3\pi}K_0
\left(\sqrt{1-x}\,Q\,b\right)
\label{hattheq}
\eeq
where $K_0$ is the modified Bessel function of order zero. The Sudakov exponent
$S$ in (\ref{fpgeq}), comprising those gluonic radiative corrections
(in next-to-leading-log approximation) not taken 
into account in the evolution of the wave function, is given by
\beq
S(x,b,Q)=s(x,b,Q)+s(1-x,b,Q)-\frac{4}{\beta_0}
\ln\frac{\ln\left(\mu/\Lambda_{QCD}\right)}{\ln\left(1/b\,\Lambda_{QCD}\right)}
\label{sudeq}
\eeq
where a Sudakov function $s$ appears for each quark line entering the hard
scattering amplitude. The last term in (\ref{sudeq}) arises from the
application of the renormalization group equation ($\beta_0=11-\frac{2}{3}
n_f$). A value of $200$ MeV for $\Lambda_{QCD}$ is used and $\mu$ is taken to 
be the largest mass scale appearing in the hard scattering amplitude, i.e.
$\mu=\max\left(\sqrt{1-x}\,Q,1/b\right)$. For small $b$ there is no suppression
from the Sudakov factor; as $b$ increases the Sudakov factor decreases,
reaching zero at $b=1/\Lambda_{QCD}$. For even larger $b$ the Sudakov is set 
to zero. The Sudakov function $s$ has been calculated by Botts and Sterman
\cite{BoSte:89} using resummation techniques; its explicit form can be found
in \cite{DaJaKro:95}. Due to the properties of the Sudakov factor any 
contribution is damped asymptotically, i.e. for $\ln (Q^2/\mu_0^2)\to\infty$, 
except those from configurations with small quark-antiquark separations and, 
as can be shown, the limiting behaviour $F_{\pi\gamma}\to\sqrt 2 f_\pi/Q^2$ 
emerges, a result which as been derived previously \cite{BroLe:81,WalZer:72}. 
$b$ plays the role of an infrared cut-off; it sets up the interface between 
non-perturbative soft gluon contributions - still contained in the hadronic 
wave function - and perturbative soft gluon contributions accounted for by the 
Sudakov factor. Hence, the factorization scale $\mu_F$ is to be taken as $1/b$.

Employing the asymptotic wave function, (\ref{wfpareq}), (\ref{gausskteq}) and
$\phi_{AS}$, we find, from (\ref{fpgeq}), the numerical results for the 
\pgtff\ displayed in Fig.~1. Obviously there is very good agreement with the 
data \cite{Sav:95,Beh:91} above $Q^2\simeq 1$ GeV$^2$. At $8$ GeV$^2$ about 
$85$\% of the asymptotic value has been reached. 
We emphasize that there is no free parameter in our 
approach to be fitted to the data once the wave function is chosen and the 
transverse size parameter is fixed through (\ref{p0ggeq}). For comparison we 
also show in Fig.~1 results obtained with the CZ wave 
function, (\ref{wfpareq}), (\ref{gausskteq}) and $\phi_{CZ}$ ($B_2=2/3$, 
$B_n=0$ for $n>2$ in (\ref{evoleq})). That prediction overshoots the data 
markedly. Of course, the experimental errors allow slight modifications of the 
asymptotic wave function. In order to give a quantitative estimate of the 
allowed modifications we fit the expansion coefficient $B_2$ to the data 
assuming $B_n=0$ for $n>2$ and choosing again $\mu_0=0.5$ GeV. For each value 
of $B_2$ the transverse size parameter $a$ is fixed through (\ref{p0ggeq}). A 
best fit to the data above $1$ GeV$^2$ provides $B_2^{mpa}=-0.006\pm0.014$ (with 
$a=864$ MeV), i.e. a value compatible with zero. In \cite{BroHuaLe:83} a 
modification of the asymptotic wave function is proposed where $\phi_{AS}$ is 
multiplied by the exponential $\exp\left[-m_q^2a^2/x(1-x)\right]$. The 
parameter $m_q$ represents a constituent quark mass of, say, $330$ MeV. A 
similar wave function is constructed by Dorokhov \cite{Dor:95} from the 
helicity and flavour changing instanton force. Although wave functions of this 
type contradict the constraint i) derived by Chibisov and Zhitnitsky 
\cite{ChiZhi:95}, they cannot be excluded absolutely since the constraint i) 
is obtained under a duality assumption the validity of which is not 
guaranteed. In any case we have convinced ourselves that the wave function 
given in \cite{BroHuaLe:83} provides similarly good results for the \pgff\ as 
the asymptotic wave function itself. This is not a surprise since both the 
wave functions differ from each other only in the end-point regions. 
Contributions from these regions are strongly suppressed by the Sudakov factor.

It is instructive to compare the leading twist result for the \pgff\ 
\cite{BroLe:80,BroLe:81}
\beq
\fpg{Q^2}=\frac{\sqrt 2}{3}\,\langle x^{-1}\rangle\frac{f_\pi}{Q^2}\;
               [\,1\,+\,\frac{\alpha_s(\mu_R)}{2\pi}
               K(Q,\mu_R)\,+\,{\cal O}(\alpha_s^2)\,]
\label{leadtwisteq}
\eeq
with the data above $Q^2\simeq 3$ GeV$^2$ which, within errors, are just 
compatible with a $Q^2$-dependence according to dimensional
counting. $\mu_R$ represents the renormalization scale. The factor $K$
depends on the distribution amplitude. Using 
the expansion (\ref{evoleq}), one finds for the $x^{-1}$ moment of the \da
\beq
\langle x^{-1}\rangle=3\left[1+B_2\left(\frac{\alpha_s(\mu_F)}{\alpha_s(\mu_0)}
\right)^{50/81}+...\right].
\label{leadmomeq}
\eeq
Neglecting again terms with $n>2$ and also the $\alpha_s$ corrections
in (\ref{leadtwisteq}), we obtain a value of $-0.39\pm0.05$ for 
$B_2^{LO}$ from a fit to the data ($\mu_F=Q$ in this case). The face
value of $B_2^{LO}$ corresponds to $\langle x^{-1}\rangle^{LO}=2.39$ 
(at $Q^2=8$ GeV$^2$) which is to be contrasted with the values of $3$ 
and $4.01$ (at $8$ GeV$^2$) for the asymptotic and the CZ 
distribution amplitude respectively. 

Braaten \cite{bra:83} has calculated the $\alpha_s$ corrections (in
the $\overline{MS}$ scheme) in (\ref{leadtwisteq}). His analysis is however
incomplete in so far as only the $\alpha_s$ corrections to the hard
scattering amplitude have been considered but the corresponding
corrections to the kernel of the evolution equation for the pion's
distribution amplitude were ignored. As has been shown by M\"uller
\cite{mue:95} recently in next-to-leading order the evolution provides
logarithmic modifications in the end-point regions for any
distribution amplitude, i.~e.~for the asymptotic one too.
An estimate however reveals that the modifications of the evolution
behaviour in next-lo-leading order are very small for the asymptotic
distribution amplitude ($\alpha_s$ evaluated in two-loop with 
$\Lambda^{\overline{MS}}_{n_F=3} = 200\,{\rm MeV}$), and can safely be
neglected here. For the CZ distribution amplitude
these effects seem to be somewhat larger than for the asymptotic one
but still the total $\alpha_s$ corrections are dominated by those to
the hard scattering amplitude. The $\alpha_s$ corrections amount to 
$-10\%$ in the case of the CZ distribution amplitude.  
Hence, also in the leading twist analysis in next-to-leading order the
CZ \da\ is clearly at variance with the data. 
Next we want to determine the expansion coefficient $B_2$ in the
next-to-leading order leading twist analysis in order to quantify the
deviations from the asymptotic distribution amplitude required by the
$F_{\pi\gamma}$ data. For this purpose we evaluate the $K$ factor in
(\ref{leadtwisteq}) from the expressions for the $\alpha_s$
corrections given in \cite{bra:83} ($\mu_R=\mu_F=Q$) and neglect the 
modifications of the evolutions in
next-to-leading order. From a fit of (\ref{leadtwisteq}) to the form
factor data we find $B_2^{NLO}=-0.17\pm0.05$ corresponding to 
$\langle x^{-1}\rangle^{NLO}=2.74$ at $Q^2=8$ GeV$^2$. According to
what we said above such small a value of $B_2$ will not be altered
substantially by the modifications of the evolution behaviour to that order.
Thus, the leading twist analysis requires a distribution amplitude
which is a little narrower than the asymptotic one. In the modified
perturbative approach, on the other hand, the asymptotic wave function
works well since the QCD corrections condensed in the Sudakov factor,
and the transverse degrees of freedom provide the required
$Q^2$-dependent suppressions. It is to be stressed that the Sudakov
factor already takes into account the leading and next-to-leading logs
of the $\alpha_s$ corrections.

Other models, applicable at large as well as at low $Q^2$, provide a 
parameterization of the
form factor as
\beq
\fpg{Q^2}=A/(1+Q^2/s_0).
\label{interpoleq}
\eeq
Thus, Brodsky and Lepage \cite{BroLe:81} propose that parameterization as an
interpolation between the two limits, $\fpg{Q^2=0}=A=(2\sqrt 2\pi^2f_\pi)^{-1}$
known from current algebra and the limiting behaviour $\sqrt 2 f_\pi Q^{-2}$.
Hence, $s_0=4\pi^2f_\pi^2=0.67$ GeV$^2$ in that model. The interpolation 
formula works rather nicely, its $Q^2$-dependence is similar to that one 
predicted by the modified perturbative approach. The vector meson dominance 
model leads to (\ref{interpoleq}) under the neglect of contributions from the 
$\phi$ meson and by ignoring the small mass difference between the $\rho$ and 
$\omega$ meson. The constant $A$ is related to
\beq
A=\frac{g_{\pi\rho\gamma}}{f_\rho}+\frac{g_{\pi\omega\gamma}}{f_\omega}
\label{vmdnormeq}
\eeq
in the vector meson dominance model. $s_0$ equals $m_\rho^2$ where $m_\rho$ is 
the $\rho$ meson mass. Inserting the known values of coupling constants 
\cite{Dum:83}, one finds for $A$ a value of $0.269\pm0.019$ in agreement with 
the current algebra value. The vector meson dominance model is in accord with 
the present data although its asymptotic value ($A\,m_\rho^2\simeq0.16$) 
differs from our one. A QCD sum rule analysis \cite{rad:95} also
provides results similar to (\ref{interpoleq}). In order to discriminate among 
the various models data extending to larger values of momentum transfer 
and/or with smaller errors as the present ones are needed.

Let us now turn to the discussion of the implications of our findings, namely 
that the perturbative analysis of the \pgtff\ requires the asymptotic pion 
wave function and apparently excludes strongly end-point concentrated wave 
functions like the one proposed by Chernyak and Zhitnitsky. Since the wave 
functions are universal, process-independent objects they should also be used 
in other large momentum transfer exclusive reactions involving pions, as for 
example the electromagnetic form factor of the pion or 
$\gamma\gamma\to\pi\pi$. As is well-known the leading twist results are only 
in agreement with experiment provided an end-point concentrated wave function 
respectively distribution amplitude is utilized\footnote{
Taking the moment $\langle x^{-1}\rangle$ which represents a soft,
process-independent parameter, from our leading twist analysis of
$F_{\pi\gamma}$, we find  a value for the pion form factor too small as compared
to the admittedly poor data (note: $F_{\pi} \sim \langle x^{-1}\rangle^2$) 
even when $\alpha_s$ corrections are considered \cite{fie:81}.}. 
This apparent agreement with experiment is, as we already 
mentioned, only obtained at the 
expense of strong contributions from the soft end-point regions where the use 
of perturbation theory is unjustified. This is to be contrasted with the 
modified perturbative approach where the end-point regions are strongly 
suppressed and a theoretically self-consistent perturbative contribution is 
obtained. However, as shown in \cite{JaKro:93} for the case of the \pff, the 
perturbative contributions evaluated with both the wave functions, the 
asymptotic one and the CZ one, are too small as compared with 
the data. At this point we remind the reader of the fact that the pion form 
factor also gets contributions from the overlap of the initial and final state 
soft wave functions $\hat\Psi_0$ (\ref{wfpareq}). Formally the perturbative 
contribution to the pion form factor represents the overlap of the large 
momentum tails of the wave functions while the overlap of the soft parts of 
the wave functions is customarily assumed to be negligible at large $Q$. 
Examining the validity of that presumption by estimating the Feynman 
contribution from the asymptotic wave function, one finds results of 
appropriate magnitude to fill in the gap between the perturbative contribution 
and the data of Ref.~\cite{Beb:76}. The results exhibit a broad flat maximum 
which, for momentum transfers between $3$ and about $15$ GeV$^2$, simulates 
the dimensional counting behaviour\footnote{
At large $Q^2$ the Feynman contribution is suppressed by $1/Q^2$ as compared 
to the perturbative contribution. The latter dominates the elastic form factor 
only for $Q^2\gsim50$ GeV$^2$. This value of $Q^2$ is, however, very sensitive 
to the end-point behaviour of the wave function, little modifications may 
change it considerably.}. For the CZ wave function, on the 
other hand, the Feynman contribution exceeds the data significantly. Similar 
large Feynman contributions have also been obtained by other authors 
\cite{IsgLle:89,KisWan:93,Dor:95}. Thus, the small size of the perturbative 
contribution to the elastic form factor finds a comforting although 
model-dependent explanation, a fact which has been pointed out by
Isgur and Llewellyn Smith \cite{IsgLle:89} long time ago.

The structure function of the pion offers another possibility to test the wave 
function against data. As has been shown in \cite{BroHuaLe:83} the parton 
distribution functions are determined by the Fock state wave functions. Since 
each Fock state contributes through the modulus squared of its wave function 
integrated over transverse momenta up to $Q$ and over all fractions $x$ except 
those pertaining to the type of parton considered, the contribution from the 
valence Fock state should not exceed the data of the valence quark structure 
function. As discussed in \cite{JaKroRau:94,HuaMaShe:94} the asymptotic wave 
function respects this inequality while the CZ one again 
fails dramatically.

To conclude the asymptotic pion wave function, respecting all theoretical 
constraints, provides a consistent and theoretically satisfying description of 
the \pg\ and the pion's electromagnetic form factor and is compatible with the 
pion's valence quark distribution function. The pion's electromagnetic form 
factor is controlled by soft physics (which can be modelled as the
Feynman contribution for the asymptotic wave function) in the experimentally 
accessible range of momentum transfer in contrast to the \pgtff\ which is 
dominated by the perturbative contribution. We note that similar 
observations about the smallness of the perturbative contributions and the 
dominance of the Feynman contributions have been made in the case of the 
nucleon's form factor \cite{BoKro:95}. Of course, the present quality of 
the data does not allow to pin down the form of the wave function exactly. 
Little modifications of the asymptotic wave function can not be excluded as 
yet. On the other hand, the CZ wave function is in conflict 
with the data and ought to be discarded. This is also true for other strongly 
end-point concentrated wave functions. The use of such wave functions in the 
analyses of other exclusive reactions involving pions, e.g. 
$\gamma\gamma\to\pi\pi$ or $B\to\pi\pi$, seems to be unjustified (if one 
accepts the process-independence of the wave function) and likely leads to 
overestimates of the perturbative contributions. The next-to-leading
order leading twist analysis of the $F_{\pi\gamma}$ form factor
(possible for $Q^2 \leq 3\,{\rm GeV}^2$) also reveals that the wave
function or better the distribution amplitude in that case, is close
to the asymptotic one but a little narrower than it. This result is to
be contrasted with the modified perturbative approach where the
asymptotic wave function works well; the required $Q^2$-dependent
suppression is provided by the Sudakov factor and the transverse
degrees of freedom. In any case a systematic next-to-leading order
analysis of exclusive reactions involving pions is required.

\newpage
{\parindent0cm\Large\bf{Figure captions}}\\[1em]

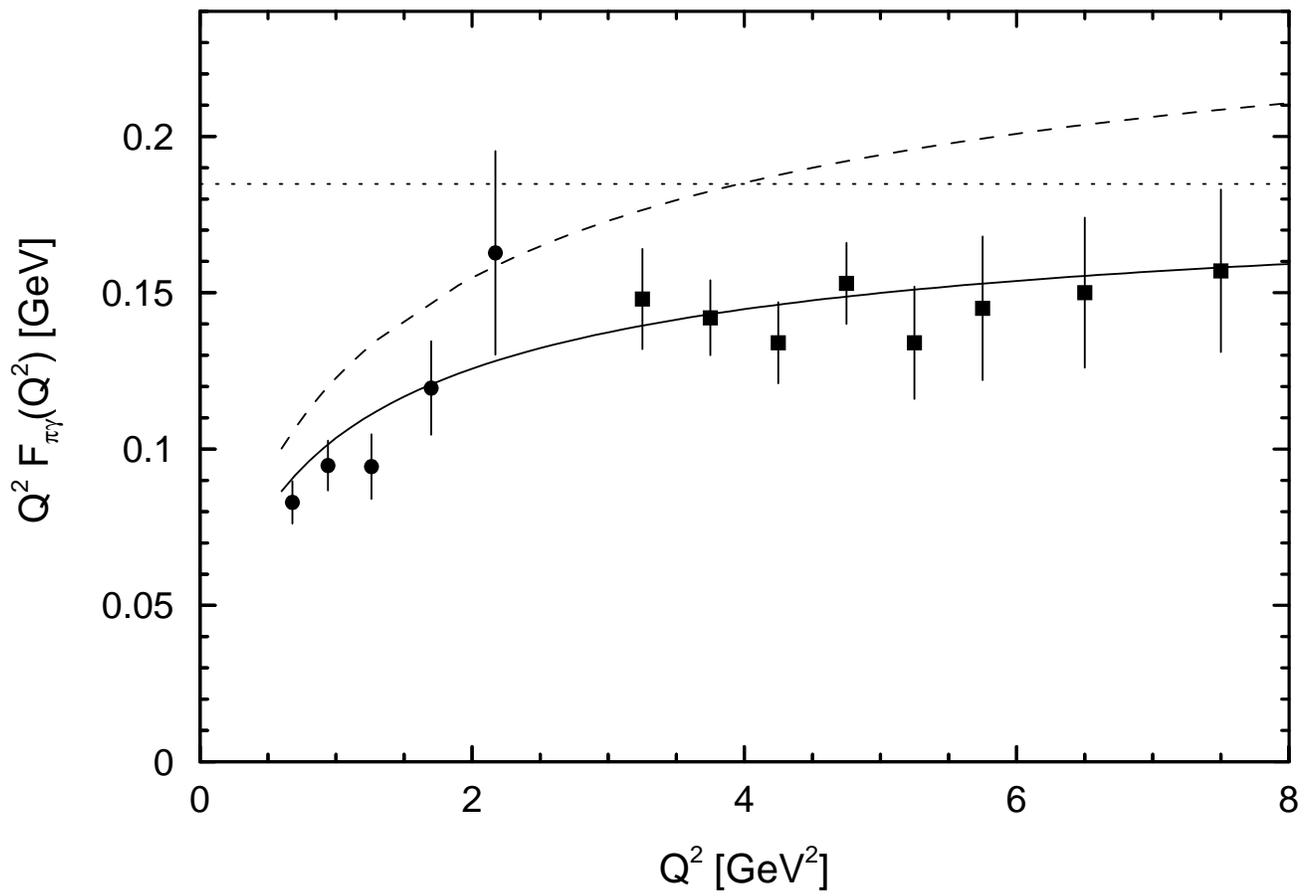
Fig.~1. The scaled \pgtff\ vs. $Q^2$. The solid (dashed) line represents 
the results obtained with the modified perturbative approach using the
asymptotic (CZ) wave function. The evolution of the CZ wave function is taken into
account. The dotted line represents the limiting behaviour 
$\sqrt 2f_\pi$. Data are taken from \cite{Sav:95,Beh:91}.\\[1em]

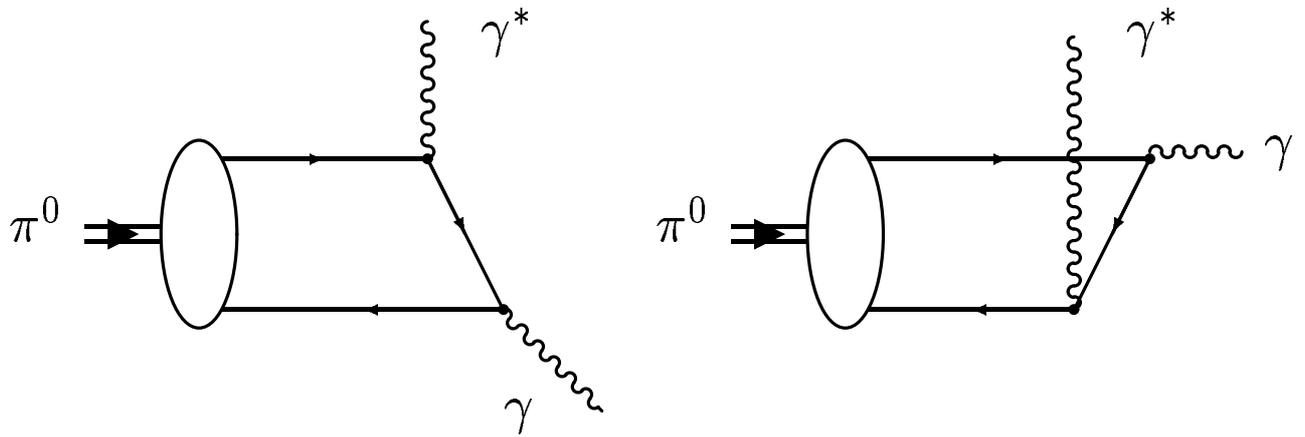
Fig.~2. The basic graphs for the \pgtff.


\newpage

\begin{thebibliography}{99}

\bibitem{CheZhi:82}{V.~L.~Chernyak and A.~R.~Zhitnitsky,
Nucl.~Phys.~B201 (1982) 492.}

\bibitem{Beb:76}{C.~J.~Bebek et al.,
Phys.~Rev.~D13 (1976) 25 and D17 (1978) 1693.}

\bibitem{IsgLle:89}{N.~Isgur and C.~H.~Llewellyn Smith,
Nucl.~Phys.~B317 (1989) 526.}

\bibitem{Rad:91}{A.~V.~Radyushkin,
Nucl.~Phys.~A532 (1991) 141c and references therein.}

\bibitem{JaKro:93}{R.~Jakob and P.~Kroll,
Phys.~Lett.~B315 (1993) 463; B319 (1993) 545(E).}

\bibitem{KisWan:93}{L.~S.~Kisslinger and S.~W.~Wang,
Nucl.~Phys.~B399 (1993) 63;\\
P.~L.~Chung, F.~Coester and W.~N.~Polyzou, 
Phys.~Lett.~B205 (1988) 545;\\
V.~Braun, I.~Halperin, 
Phys.~Lett.~B328 (1994) 457.}

\bibitem{Dor:95}{A.~E.~Dorokhov, 
preprint IFKP-TH 36/95, Pisa (1995).}

\bibitem{Sav:95}{CLEO coll., V.~Savinov et al.,
proceedings of the PHOTON95 workshop, Sheffield (1995),
eds. D.~J.~Miller et al., World Scientific.}

\bibitem{Beh:91}{CELLO coll., H.-J.~Behrend et al.,
Z.~Phys.~C49 (1991) 401.}

\bibitem{bra:83} {E.~Braaten, Phys.~Rev.~D28 (1983) 524.}

\bibitem{Gor:89}{A.~S.~Gorski\u\i,
Sov.~J.~Nucl.~Phys.~50 (1989) 498.}

\bibitem{JaKroRau:94}{R.~Jakob, P.~Kroll and M.~Raulfs,
Jour.~of Phys.~G22 (1996) 45; P.~Kroll, proceedings of the 
PHOTON95 workshop, Sheffield (1995),
eds. D.~J.~Miller et al., World Scientific.} 

\bibitem{Ong:95}{S.~Ong, 
Phys.~Rev.~D52 (1995) 3111.}

\bibitem{BroHuaLe:83}{S.~J.~Brodsky, T.~Huang and G.~Lepage,
Banff Summer Institute, Particles and Fields 2, p. 142, A.~Z.~Capri and
A.~N.~Kamal (eds.), 1983.}

\bibitem{BroLe:80}{G.~P.~Lepage and S.~J.~Brodsky,
Phys.~Rev.~D22 (1980) 2157.}

\bibitem{BoSte:89}{J.~Botts and G.~Sterman,
Nucl.~Phys.~B325 (1989) 62.}

\bibitem{ChiZhi:95}{B.~Chibisov and A.~R.~Zhitnitsky,
Phys.~Rev.~D52 (1995) 5273.}

\bibitem{mue:95}{D.~M\"uller, Phys.~Rev.~D51 (1995) 3855.}

\bibitem{LiSte:92}{H.~N.~Li and G.~Sterman,
Nucl.~Phys.~B381 (1992) 129.}

\bibitem{BoJaKro:94}{J.~Bolz, R.~Jakob, P.~Kroll, M.~Bergmann and
N.~G.~Stefanis,
Z.~Phys.~C66 (1995) 267 and Phys.~Lett.~B342 (1995) 345.}

\bibitem{fie:81}{R.~D.~Field et al., Nucl.~Phys.~B186 (1981) 29; 
F.~M.~Dittes and A.~V.~Radyushkin, Sov.~J.~Nucl.~Phys.~34 (1981) 29;
E.~Braaten and S.-M.~Tse, Phys.~Rev.~D35 (1987) 2255.}

\bibitem{DaJaKro:95}{M.~Dahm, R.~Jakob and P.~Kroll,
Z.~Phys.~C68 (1995) 595.}

\bibitem{BroLe:81}{S.~J.~Brodsky and G.~P.~Lepage,
Phys.~Rev.~D24 (1981) 1808.}

\bibitem{WalZer:72}{T.~F.~Walsh and P.~Zerwas,
Nucl.~Phys.~B41 (1972) 551.}

\bibitem{Dum:83}{O.~Dumbrajs, R.~Koch, H.~Pilkuhn, G.~C.~Oades, H.~Behrens, 
J.~J.~de~Swart and P.~Kroll, 
Nucl.~Phys.~B216 (1983) 277.}
 
\bibitem{rad:95} {A.~V.~Radyushkin, R.~T.~Ruskov, preprint
      hep-ph/9511270 (1995).} 
 
\bibitem{HuaMaShe:94}{T.~Huang, B.-Q.~Ma and Q.-X.~Sheng, 
Phys.~Rev.~D49 (1994) 1490.}

\bibitem{BoKro:95}{J.~Bolz and P.~Kroll, 
preprint WU-B 95-35, Wuppertal (1995).}

\end{thebibliography}
\end{document}